\documentclass[journal,transmag]{IEEEtran}
\def\rd{{\rm d}}

\def\vx{{\bf x}}
\def\vz{{\bf z}}
\def\vnu{\mbox{\boldmath$\nu$}}

\ifCLASSINFOpdf
\else
\fi
\usepackage{amsmath}
\usepackage{amssymb}
\usepackage{mathrsfs}
\usepackage{chemarrow}
\begin{document}
%
\title{Nonlinear Stochastic Dynamics of Complex Systems, III:
Noneqilibrium Thermodynamics of
Self-Replication Kinetics}


\author{\IEEEauthorblockN{David B. Saakian\IEEEauthorrefmark{1,}\IEEEauthorrefmark{2} and
Hong Qian\IEEEauthorrefmark{3}}
\IEEEauthorblockA{\IEEEauthorrefmark{1}Institute of Physics, 
Academia Sinica, Nankang, Taipei 11529, Taiwan
}
\IEEEauthorblockA{\IEEEauthorrefmark{2}A. I. Alikhanyan 
National Science Laboratory (Yerevan Physics Institute) Foundation\\ 
2 Alikhanian Brothers Street, Yerevan 375036, Armenia
}
\IEEEauthorblockA{\IEEEauthorrefmark{3}Department of Applied Mathematics, University of Washington, Seattle,
WA 98195-3925, USA}
\thanks{
Corresponding authors: D. B. Saakian (email: saakian@phys.sinica.edu.tw) 
and H. Qian (email: hqian@u.washington.edu).}}

\markboth{}%
{Shell \MakeLowercase{\textit{et al.}}: Bare Demo of IEEEtran.cls for IEEE Transactions on Magnetics Journals}
%



\IEEEtitleabstractindextext{%
\begin{abstract}
We briefly review the recently developed, Markov process based
isothermal chemical thermodynamics for nonlinear,
driven mesoscopic kinetic systems.  Both the instantaneous 
Shannon entropy {\boldmath $S[p_{\alpha}(t)]$} and 
relative entropy {\boldmath $F[p_{\alpha}(t)]$}, defined 
based on probability distribution {\boldmath $\{p_{\alpha}(t)\}$}, 
play prominent roles.  The theory is general;
and as a special case when a chemical 
reaction system is situated in an equilibrium environment, it agrees perfectly with Gibbsian chemical thermodynamics:
{\boldmath $k_BS$} and {\boldmath $k_BTF$} become thermodynamic 
entropy and free energy, respectively.   We apply this 
theory to a fully reversible autocatalytic reaction kinetics,
represented by a Delbr\"{u}ck-Gillespie process, in a 
chemostatic nonequilibrium environment. The open, driven chemical 
system serves as an archetype for biochemical self-replication.
The significance of {\em thermodynamically consistent} 
kinetic coarse-graining is emphasized.  In a kinetic system where death of a biological organism is treated as the reversal of its birth, the meaning of mathematically emergent ``dissipation'', which is not related to the heat measured in terms of {\boldmath $k_BT$},  remains to be further investigated.

\end{abstract}

\begin{IEEEkeywords}
Biophysics, chemical master eqution, chemical potential, detailed balance, 
entropy, Gillespie algorithm, information.
\end{IEEEkeywords}}

\maketitle

\IEEEdisplaynontitleabstractindextext

%
\IEEEpeerreviewmaketitle

\section{Introduction}
%
%
%
%
\IEEEPARstart{I}{n} Part I and Part II of this series \cite{part-1,part-2}, a chemical reaction kinetic perspective on complex systems
in terms of a mesoscopic stochastic nonlinear kinetic
approach, e.g., Delbr\"{u}ck-Gillespie processes, as well as a
stochastic nonequilibrium thermodynamics (stoc-NET) in
phase space, have been presented.  Part I provides
an overview. Part II, motivated by both
Gibbsian statistical mechanics and information
theory, focuses on a parametric family of probability
distributions $p(x,\beta)$ as a novel {\em theoretical
device} for exploring a single nonequilibrium steady state
probability $P(x)$, $x\in\mathscr{S}$, where
$\mathscr{S}$ is a denumerable set of events, and
$(\mathscr{S},\mathcal{F},P)$ is the underlying
probability space.

    The relation between Gibbsian statistical mechanics
and information theory \cite{sw-book}
has generated endless discussions ranging from ``they are
completely unrelated except sharing a common term
`entropy','' to ``they are parts of a single unifying maximum
entropy principle''.  Throwing a monkey wrench into the
midst, we have articulated in \cite{part-2} the
notion that minus-log-$p^{ss}$ can be a very legitimate
potential function of an entropic force in a complex dynamics.
The origin of this practice was traced back to equilibrium statistical
chemical thermodynamics with an exact coarse-graining, and
J. G. Kirkwood's concept of {\em potential of mean force}.
The seemingly naive question still remains:
What is information?  Is it a state of a complex
system, or a character of its dynamics, or a form of ``energy''
\cite{sano-10,jarzynski-12}?  In the engineering specific
context first discussed by Shannon,
information is the mathematical symbols that represent certain
messages in communication.  Therefore, the theory of
information is a theory concerning coding, representations,
and communications under uncertainty.   On the other hand,
in a much broader sense, particularly in the early days of
applying information theory to biology, mathematical
information theory was used essentially as an applied
probability with a singular emphasis on the notion of
Shannon entropy \cite{johnson-1970,schneider}.

Sidestepping these issues, we
follow Kolmogorov's axiomatic theory of probability in which
information is simply
a very particular function of a random variable $\vx$
with probability distribution $p_{\vx}$ 
\cite{kolmogorov-68,qian_pre_02}:
$H_{\vx}(x)=-\ln p_{\vx}(x)$, $x\in\mathscr{S}$,
whose expected value gives Gibbs-Shannon formula.

\begin{subequations}
\label{equation001}
It is important to remember that in
Gibbs' statistical ensemble theory of equilibrium
matters \cite{fowler-book}, the
probability of a state $\vx$ of a matter that is in contact
with a temperature bath is
\begin{equation}
             p_{\vx} = \frac{e^{-E_{\vx}/k_BT} }{\Xi(T)},
       \  \textrm{ where } \
           \Xi(T) = \sum_{\vx\in\mathscr{S}}
               e^{-E_{\vx}/k_BT},
\end{equation}
in which $E_{\vx}$ is the internal energy of the state $\vx$,
$T$ is temperature in Kelvin, $k_B=1.38064852
\times 10^{-23}\ m^2\ kg\ s^{-2}\ K^{-1}$
is Boltzmann's constant, $\mathscr{S}$ is the set
containing all possible states of the matter.  The theory
of statistical mechanics states that the entropy of the
matter is then
\begin{equation}
     S(T) = -k_B\sum_{\vx\in\mathscr{S}} p_{\vx}\ln p_{\vx}
       = \frac{\partial}{\partial T}\Big( k_BT\ln\Xi(T)\Big).
\end{equation}
In statistical thermodynamics, $F(T)\equiv -k_BT\ln\Xi(T)$ is
called free energy, and entropy is $-\partial F/\partial T$.  One
also has the important equation
\begin{equation}
                      F=\overline{E}-TS,
\end{equation}
in which $\overline{E}$ is the mean internal energy
\begin{equation}
        \overline{E}(T) = \sum_{\vx\in\mathscr{S}} E_{x} p_{\vx}.
\end{equation}
\end{subequations}
Viewed from these four widely known mathematical equations, the
{\em theoretical device} discussed in \cite{part-2}
simply provides an extended exploration of the information
content of $P(x)$ and its underlying $(\mathscr{S},\mathcal{F},P)$.

    There is a growing interest in quantitative
relations between the theory of thermodynamics that
rules many aspects of the biochemical world and the biological
dynamics of organisms, their ecology and evolution.
``Every species of living thing can make a copy of itself
by exchanging energy and matter with its surroundings.''
\cite{jle-1} A living organism has two distinct aspects:
One with ``genomic information'' which is carried
in the sequence of polynucleic acids \cite{schneider},
possibly including methylation modifications of base-pairs; and
the other is a space-time biochemical dynamics in terms of
molecules, which can exhibit chemical, mechanical, and
electrical characteristics. The time scales of these
two aspects are vastly different:  The former cumulates
changes generation after generation, which constitutes an
evolutionary process.  The latter defines ``what is living''
for each and every individual organism.  The relationship
between these two has been the subject of many scholarly
studies in the field of biology 
\cite{roughgarden,stadler-stadler,kirschner-gerhart,noble,weinberg-14}.

  To a classic physicist or chemist, the cellular biochemical
processes are various transformations of energy and
matters; to a modern cell biologist, the same processes are
usually described as regulation and processing
of {\em informations}.  The physical and chemical processes
are interpreted with biological functions \cite{hhlm}.
The present paper is an attempt to
provide a treatment of this latter aspect in terms of
a stochastic kinetic theory.  In particular, we analyze
rigorously a chemical model that makes the notion of
``making a copy of itself by exchanging energy and matter
with its surroundings'' precise and quantitative.
To this end, we revisit a kinetic model for a fully
reversible chemical reaction system that consists of
an autocatalytic step $A + X\longrightarrow 2X$
\cite{vellela-qian,bishop-qian}. This model first appeared
in the wonderful book
by the late Professor Keizer \cite{keizer-book}.
This extremely simple model, however, offers us an
insight on why a truly meaningful
chemical thermodynamic study of biological
replication is premature:  This is because neither
the current cosmological time scale of our
universe nor the size of entire biological population
on our planet are sufficiently large to provide a
meaningful estimation of the probability of
``two [daughter cells] somehow spontaneously revert back
into one'' \cite{jle-1}.  Note the ``revert back'' should
not be a fusion of two cells, rather it has to
be a process that reverses the entire event-by-event
of a cell replication.

It is the non-zero
probability of such completely inconceivable processes,
in a physicist's world, that provides an estimation of thermodynamic
heat production.  When a biologist considers only feasible
events to a living organism being birth and death, the quantitative connection to the world of thermal physics has already been
lost.\footnote{On the other hand,
the upper bound of such an extremely small probability 
can be estimated from knowing the minimal amount of 
calories required to reproduce an organism using the
very equation (2) in \cite{jle-1}.
} 
This difficulty has been hinted in \cite{ruelle-ejp}:
``If we assume that the process $I\rightarrow II$ is
`irreversible' this implies that the time-reversed dynamics is very
unstable so that $\Pr\{II\rightarrow I\}$ is hard to estimate
precisely. One uses instead a bigger probability
$\Pr\{II\rightarrow I\}$ based on observable processes.''
Unfortunately, any remotely reasonable substitution for
an authentic de-birth is currently out of reach; death is not an
acceptable alternative even for a lower bound estimation.

    The paper is structured as follows:  In Sec. \ref{sec:msnt}
we give a brief review of the modern theory of stochastic
nonequilibrium thermodynamics of Markov systems
\cite{ge-qian-pre-10,qkkb-jpcm-16,ge-qian-2016,qian-epjst-15}, as
presented in Part I \cite{part-1}. While many of the terminologies
and mathematical relations can be found in the review articles
\cite{jarzynski,seifert,vdb-e} and classic text \cite{dm-book}, we
like to point out that the coherent narrative constructed based on
the mathematics, e.g., their interpretations, is unique.  We
shall show how mathematical logic dictates that 
entropy in systems with uniform stationary
probability should be replaced by a relative entropy, also known
as free energy, in systems with non-uniform stationary
probability, and how entropy production rate $e_p$ arises as a
distinctly different concept as entropy change $\frac{\rd S}{\rd
t}$.  Then in Sec. \ref{sec:ocm}, we first give the standard
chemical thermodynamic treatment of the simple kinetic system with
two reversible reactions
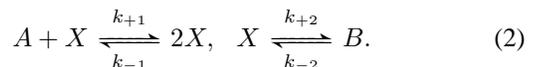
\begin{equation}
   A + X  \ \underset{k_{-1}}{\overset{k_{+1}}{\rightleftharpoonsfill{24pt}}}   \  2X,  \   \
     X  \ \underset{k_{-2}}{\overset{k_{+2}}{\rightleftharpoonsfill{24pt}}}   \  B.
\label{the-rxn}
\end{equation}
Note, if one neglects the two ``backward'' reactions,
e.g., letting
 $k_{-1}=k_{-2}=0$, then the chemical kinetics
can be described by the differential equation
\begin{equation}
    \frac{\rd x}{\rd t} = g x - k_{+2} x,  \  \
            g\equiv k_{+1}a.
\label{eq-002}
\end{equation}
Eq. \ref{eq-002} describes the
``birth and death'' of individual $X$ molecules.

The simple chemical model in (\ref{the-rxn}) has a clear
self-replication characteristics, it allows a rigorous
thermodynamics analysis of the replication/synthesis aspect of a
biological organism.  Certainly the individual $X$ lacks other
fundamentals of a living being:  An individual $X$ itself is a
dead molecule, not a living organism. Therefore, while it is
meaningful to ask the heat dissipation in a self-replication from
a complex chemical synthesis standpoint \cite{kjelstrup}, it
might not be a sufficient model for the heat dissipation in the
self-replication of a living organism since even just being alive,
an organism has a continuous heat dissipation as an individual
entity, e.g., active metabolism.  Simply put: a living organism,
which has both {\em metabolism} and {\em self-replication},
already has a basal level of heat dissipation due to the former
even in the absence of the latter.  The present analysis, however,
makes a conceptual separation between the two processes in an
organism.  We note that the open chemical systems theory of motor
protein chemomechanics serves as a concrete model for the former
\cite{kolomeisky-book}.

In Sec. \ref{sec:dgp}, a mesoscopic stochastic
thermodynamic treatment of the replication kinetics is
carried out.  Entropy production rate for the kinetic
process is studied.  Then in Sec. \ref{sec:KandT},
we articulate a distinction between thermodynamically
consistent and inconsistent kinetic approximations.
The paper concludes with a discussion in Sec. \ref{sec:dis}.

\section{A Modern Stochastic Nonequilibrium Thermodynamics
of Markov Systems}
\label{sec:msnt}

    The very brief summary of Gibbs' equilibrium statistical
thermodynamics, given in Eq. \ref{equation001}a-d,
illustrates the universality of the
entropy function, be it in information theory or in thermal
physics.   Indeed there is now a more
unifying theory of entropy and relative entropy, also known
as Kullback-Leibler (KL) divergence in information theory and
free energy in statistical thermodynamics, based on the
probability theory of Markov processes that describe
complex system's dynamics in phase space \cite{ge-qian-pre-10,qkkb-jpcm-16,qian-epjst-15}.\footnote{Another unifying approach
to entropy is the theory of large deviations \cite{touchette}.
A deep relation between the entropy function in large deviations
theory and the entropy function in complex dynamics 
is Sanov's theorem.  See below as well as \cite{ge-qian-2016,ge-qian-2016-2}.
}

    The first basic assumption of this stochastic
nonequilibrium thermodynamic theory is that a complex
mesoscopic dynamical system can be represented by an
irreducible continuous time Markov process with appropriate state space $\mathscr{S}$ and transition probability rates
$q_{\alpha\beta}$, $\alpha,\beta\in\mathscr{S}$,
where $q_{\alpha\beta}=0$ if and only if $q_{\beta\alpha}=0$.
Under these assumptions, it can be shown that there exists
a unique positive stationary probability $\pi_{\alpha}$ for the Markov
system in stationarity:
\begin{equation}
     \sum_{\alpha\in\mathscr{S}} \pi_{\alpha}
        q_{\alpha\beta} = 0.
\end{equation}
For a stationarity process defined by $\{\pi_{\alpha}\}$ and 
$q_{\alpha\beta}$, a further distinction between an
{\em equilibrium steady state} and a {\em nonequilibrium
steady state} (NESS) can be made (see below): 
The former has zero entropy production rate, 
and the latter has a strictly positive entropy production rate.
This distinction is determined by the set of transition probability
rates $\{q_{\alpha\beta}\}$, which necessarily satisfy
detailed balance, $\pi_{\alpha}q_{\alpha\beta}=\pi_{\beta}q_{\beta\alpha}$
$\forall\alpha,\beta\in\mathscr{S}$, 
if and only if the steady state is an equilibrium.
The theory presented below is applicable to both equilibrium
and nonequilibrium steady state, as well as a time-dependent
non-stationary process.

    For a mesoscopic system such as the discrete chemical
reactions in (\ref{the-rxn}) with state space
$\mathscr{S}$ being non-negative integers,
a second assumption is that it has
a {\em macroscopic corresponding} continuous dynamics
$x(t)$, $x\in\mathbb{R}^n$.
Let $\omega$ be the ``size parameter'' that connects the
mesoscopic system and its macroscopic limit when
$\omega\rightarrow\infty$, with $x\equiv\tfrac{\alpha}{\omega}$
being {\em number density}.  Then,  for a large class of
such Markov systems \cite{ge-qian-2016,ge-qian-2016-2},
\begin{equation}
      \frac{ \pi_{\omega\vx} }{\omega}
             \simeq e^{-\omega\varphi(\vx)}
       \rightarrow \delta \big(\vx-\vz\big),
\label{equation003}
\end{equation}
in which function $\varphi(\vx)\ge 0$ and its global minimum is at
$\vz$.  In applied mathematical theory of singular perturbation,
Eq. \ref{equation003} is known as WKB ansatz \cite{omalley-book};
in the theory of probability, it is called {\em large deviations
principle} \cite{touchette}.

    Inspired by the similarity between the expression in Eq. \ref{equation003} and Boltzmann's law, let us define
``the internal energy density of the Markov state $\alpha$'':
\begin{equation}
     E_{\alpha} = -\frac{ \ln \pi_{\alpha} }{\omega}.
\end{equation}
Then the mean energy density at time $t$:
\begin{equation}
    \overline{E}(t) = \sum_{\alpha\in\mathscr{S}}
                     p_{\alpha}(t)E_{\alpha} =
             -\omega^{-1}\sum_{\alpha\in\mathscr{S}} p_{\alpha}(t)
               \ln\pi_{\alpha},
\end{equation}
where $p_{\alpha}(t)$ is the probability of the system in state
$\alpha$ at time $t$.  Then the free energy, also known as
Massieu potential, of the Markov system at time $t$
\begin{equation}
    F\big[p_{\alpha}(t)\big] = \omega\overline{E}-S(t)
       = \sum_{\alpha\in\mathscr{S}} p_{\alpha}(t)
          \ln\left(\frac{p_{\alpha}(t) }{\pi_{\alpha}}\right).
\end{equation}
Then, concerning these quantities, one has a series of mathematical results:

{\bf\em A free energy balance equation.}
First, an equation that is valid for systems with and without
detailed balance, and for processes in stationary and in transient:
\begin{subequations}
\label{equation007}
\begin{equation}
  \frac{\rd}{\rd t}F\big[p_{\alpha}(t)\big] = \omega E_{in}\big[p_{\alpha}(t)\big]
            - e_p\big[p_{\alpha}(t)\big],
\end{equation}
in which
\begin{equation}
        E_{in}\big[p_{\alpha}\big] :=  \frac{1}{\omega}\sum_{\alpha,\beta\in\mathscr{S}}
      \Big(p_{\alpha}q_{\alpha\beta}-p_{\beta}q_{\beta\alpha}
 \Big)\ln\left( \frac{\pi_{\alpha}q_{\alpha\beta} }{
                   \pi_{\beta}q_{\beta\alpha} } \right) 
          \ge \ 0,
\end{equation}
\begin{equation}
    e_p\big[p_{\alpha}\big] := \sum_{\alpha,\beta\in\mathscr{S}}
      \Big(p_{\alpha}q_{\alpha\beta}-p_{\beta}q_{\beta\alpha}
 \Big)\ln\left( \frac{p_{\alpha}q_{\alpha\beta} }{
                   p_{\beta}q_{\beta\alpha} } \right) \
          \ge \ 0.
\end{equation}
Eq. \ref{equation007} is interpreted as a free energy
balance equation for a Markov system with $E_{in}[p_{\alpha}]$
being the instantaneous rate of input energy, a source term,
and $e_p[p_{\alpha}]$, a sink, as the instantaneous rate of energy lost,
or entropy production rate.

\end{subequations}

    {\bf\em Entropy change, free energy change, and entropy
production.} Second, an inequality that is valid for systems with 
and without detailed balance, and for processes in stationary 
and in transient:
\begin{equation}
    \frac{\rd F(t)}{\rd t} \le 0.
\end{equation}
Combined with (\ref{equation007}),  this implies
$\omega^{-1}e_p\ge E_{in}$.

    One notices that for very particular Markov systems with $q_{\alpha\beta}=q_{\beta\alpha}$
$\forall\alpha,\beta\in\mathscr{S}$, they have a uniform
$\pi_{\alpha}\equiv$ constant.  Then
$e_p(t)=\frac{\rd S}{\rd t}$.  Such systems are analogous
to {\em microcanonical ensembles}, where the entropy production
is the same as entropy increase.

    For Markov systems with detailed balance, $E_{in}(t)\equiv 0$
$\forall t$, and
\begin{subequations}
\begin{equation}
   e_p\big[p_{\alpha}\big]
        = \frac{\rd}{\rd t}S\big[p_{\alpha}(t)\big]
         - \omega  E_{ex}\big[p_{\alpha}(t)\big] \ge 0,
\label{equation009}
\end{equation}
in which
\begin{equation}
     E_{ex}\big[p_{\alpha}\big] := \sum_{\alpha,\beta\in\mathscr{S}}
      \Big(p_{\alpha}q_{\alpha\beta}-p_{\beta}q_{\beta\alpha}
 \Big)\Big(E_{\beta}-E_{\alpha}\Big),
\end{equation}
is interpreted as the instantaneous rate of energy exchange.  The
inequality in Eq. \ref{equation009} is analogous to Clausius
inequility for spontaneous thermodynamic processes, which gives
rise to the notion of entropy production as a distinctly different
concept as $\frac{\rd S}{\rd t}$.
\end{subequations}

    {\bf\em Macroscopic limit.}
For a macroscopic system in the limit of
$\omega\rightarrow\infty$, the
probability distribution $p_{\alpha}(t)$ becomes a deterministic
dynamics $\vx(t)$, and Eq. \ref{equation007}
becomes a novel macroscopic equation \cite{ge-qian-2016,ge-qian-2016-2}
\begin{subequations}
\begin{equation}
   \frac{\rd\varphi\big[\vx(t)\big]}{\rd t}  = \textrm{cmf}\big[
              \vx(t)\big] - \sigma\big[\vx(t)\big],
\label{dphixdt}
\end{equation}
in which
\begin{eqnarray}
   \textrm{cmf}[\vx]  &=& \sum_{\ell:
          \text{all reactions} } \Big(J_{\ell}^+(\vx)-J_{\ell}^-(\vx)\Big)
\nonumber\\
      &\times &  
            \ln\left(\frac{J_{\ell}^+(\vx)}{J_{\ell}^-(\vx)}
                     e^{-\vnu_{\ell}\cdot\nabla_{\vx}\varphi(\vx)} \right),
\label{cmf}
\\
    \sigma[\vx] &=& \sum_{\ell:
          \text{all reactions} } \Big(J_{\ell}^+(\vx)-J_{\ell}^-(\vx)\Big)
               \ln\left(\frac{J_{\ell}^+(\vx)}{J_{\ell}^-(\vx)}\right),
\nonumber\\
\label{sigma}
\end{eqnarray}
$J^+_{\ell}$ and $J^-_{\ell}$ are the forward
and backward fluxes of the $\ell^{th}$ reversible reaction,
integer vector $\vnu_{\ell}$ is its stoichiometry coefficients.
cmf is the chemical motive force that sustains a reaction system
out of its equilibrium, and $\sigma$ is the macroscopic rate of
entropy production.

    If $J_{\ell}^+(\vx)=$ $J_{\ell}^-(\vx)e^{\vnu_{\ell}\cdot\nabla_{\vx}\varphi(\vx)}$
$\forall \ell, \vx$, it is known as detailed balance in chemistry, or G. N.
Lewis' law of entire equilibrium \cite{gnlewis-1925}.  In
such systems, cmf $=0$, $\varphi(\vx)$ is the Gibbs function,
$\partial\varphi(\vx)/\partial x_k$ is the chemical potential of
species $k$, and $\vnu_{\ell}\cdot\nabla_{\vx}\varphi(\vx)$ is the
chemical potential difference of the $\ell^{th}$ reaction.
\end{subequations}

    {\bf\em Macroscopic systems with fluctuations.}
For a system with very large but finite $\omega$, Eq. \ref{equation003} provides a ``universal'', asymptotic
expression for stationary, fluctuating $\vx$ \cite{touchette}:
\begin{equation}
 f_{\vx}(x) = \frac{ e^{-\omega\varphi(\vx)+\psi(\vx)} }{\Xi(\omega)}, 
\end{equation}
where
\begin{equation}
            \Xi(\omega) =  \int_{\mathbb{R}^n}
                    e^{-\omega\varphi(\vx)+\psi(\vx)}\rd\vx,
\end{equation}
in which $\varphi(\vx)$ is uniquely defined with
$\min_{\vx}\varphi(\vx) = 0$.

\section{An Open Chemical System as a Self-replicating
Entity}
\label{sec:ocm}

    We now apply the general theory in Sec. \ref{sec:msnt}
to reaction system (\ref{the-rxn}).  There are two reversible
reactions with stoichiometric coefficients $\nu_1=+1$ and $\nu_2=-1$,
respectively.  Let $x(t)$ be the concentration of $X$ at time $t$,
the kinetics of the reaction system in (\ref{the-rxn})
can be described by
\begin{eqnarray}
   \frac{\rd x(t)}{\rd t} &=& \sum_{\ell=1,2}
          \nu_{\ell}\Big(J^+_{\ell}(x)-J^-_{\ell}(x)\Big)
\nonumber\\
          &=& k_{+1}ax - k_{-1}x^2 -
                    k_{+2}x + k_{-2}b,
\label{the-ode}
\end{eqnarray}
with $J^+_1(x) = k_{+1}ax$, $J_1^-(x)=k_{-1}x^2$,
$J_2^+=k_{+2}x$, and $J_2^-=k_{-2}b$.
The first of these two reversible reactions in (\ref{the-rxn}) is
known as {\em autocatalytic}.  One can find many many examples
of this abstract system of reactions in biochemical literature;
see \cite{bishop-qian} for more extensive biochemical motivations.
One also notices that in a semi-reversible case when $k_{-2}=0$ in
the system (\ref{the-rxn}), extinction of $X$ is the only long time
fate of the kinetics.  The differential equation $\frac{\rd x}{\rd t}
= \big(g-k_{+2}\big)x-k_{-1}x^2$, however, predicts a
stable population $x^{ss} = \big(g- k_{+2}\big)/k_{-1}$.
This disagreement is known as {\em Keizer's paradox}
\cite{vellela-qian,keener}.  We shall use this model as
a concrete case of the chemical nature of
{\em exchanging of energy and matter in self-replication}.

\subsection{NESS of an open chemical
system}

    According to (\ref{the-rxn}), an $X$ molecular has transformed the
raw material in the form of $A$ into a copy of itself, a second
$X$.  The canonical mathematical description of this autocatalytic
reaction is $\frac{\rd x}{\rd t} = g x$ where the constant
$g=k_{+1}a$.  We shall use $a,x,b$ to denote the concentrations
of $A,X,B$, respectively.  Clearly, $g$ can be identified as
{\em per capita birth rate} if one is interested in the population
dynamics of $X$.  Combining with the second reaction, one
has $\frac{\rd x}{\rd t} = gx - k_{+2}x$, where $k_{+2}$
can be identified as a {\em per capita death rate}.  When
the $X$ is dead, the material in terms of atoms are in the form
of $B$.  Therefore, the ``birth'' and ``death'' of $X$, or more 
precisely the synthesis and degradation of $X$, involve
an exchange of materials, as source and waste,
with its surroundings.  One can
in fact assume that the concentrations of $A$ and $B$, as
the environment of the chemical reaction system in (\ref{the-rxn}),
are kept at constant, as a chemostat.  This is precisely why
living cells have to be ``cultured''.

	The calorie count one reads from a food label in a supermarket
gives the chemical potential difference between $A$ and $B$, 
introduced below.  This is not different from one reads
the electrical potential difference on a battery to be used for
keeping a radio ``alive''.

    One might wonder why we assume the reactions in (\ref{the-rxn})
reversible?  It turns out, as anyone familiar with chemical thermodynamics knows, one can not discuss energetics in an
irreversible reaction system.  If the $k_{-2}=0$, then the
chemical energy difference between $X$ and $B$ are
infinite; which is clearly unrealistic.    In fact, the
chemical potential difference between $A$ and $B$, in
$k_BT$ unit, is
\begin{eqnarray}
   \mu_A - \mu_B &=& \big(\mu_A-\mu_X\big) +
                  \big(\mu_X - \mu_B\big)
\nonumber\\
    &=& \ln\left(\frac{k_{+1}a}{k_{-1}x}\right) +
         \ln\left(\frac{k_{+2}x}{k_{-2}b}\right)
\nonumber\\
    &=& -\ln K^{eq}_{AB}
          +\left(\frac{a}{b}\right),
\label{dmu}
\end{eqnarray}
in which the overall equilibrium constant between
$A$ and $B$, $K^{eq}_{AB}=$
$k_{-1}k_{-2}/(k_{+1}k_{+2})$.
In a chemical or biochemical laboratory, the
equilibrium constant is usually determined in an
experiment according to
\[
            K^{eq}_{AB} = \frac{\text{equi. conc. of }A }{
                       \text{equi. conc. of }B }.
\]
If $\mu_A>\mu_B$, then there is a continuous material
flow from $A$ to $B$, even when the concentration of
$X$ is in a steady state: There is a continuous
birth and death, synthesis and degradation: metabolism
in an ``living system''.  There is an amount of
entropy being produced in the surroundings.  An agent has
to constantly generating $A$ with high chemical potential
from $B$ with low chemical potential.  This entropy
production in fact is precisely the $\mu_A-\mu_B$
\cite{ge-qian-2013}.

    The kinetics of (\ref{the-rxn}), described by Eq. \ref{the-ode}, eventually reach a steady state, with the concentration of $X$, $x^{ss}$ as the positive
root of the polynomial on the right-hand-side of (\ref{the-ode}):
\begin{equation}
    x^{ss} = \frac{1}{2k_{-1}}\left[
                 k_{+1}a-k_{+2} +\sqrt{\big( k_{+1}a-k_{+2}\big)^2
                  +4k_{-1}k_{-2}b} \right].
\end{equation}
The net flux from $A$ to $B$ in the steady state,
\begin{eqnarray}
    J^{ss}_{\text{ net } A \rightarrow B}
      &=&  J^+_1(x^{ss})-J^-_1(x^{ss})  \  =  \
       J^+_2(x^{ss})-J^-_2
\nonumber\\
        &=& \frac{1}{2k_{-1} } \left[ 2\lambda
               +k_{+2}\sqrt{\big(k_{+1}a+k_{+2}\big)^2
                  -4\lambda} \right.
\nonumber\\
        &-&  \left. \Big(ak_{+1}k_{+2}+k_{+2}^2\Big) \right],
\label{nessflux}
\end{eqnarray}
where
\[
           \lambda =ak_{+1}k_{+2}-bk_{-1}k_{-2} =
                 bk_{-1}k_{-2}\left(
                    e^{\frac{\mu_A-\mu_B}{k_BT}}-1\right).
\]
And the steady-state entropy production {\em rate} according to
Eq. \ref{sigma} is
\begin{equation}
   \sigma\big[x^{ss}\big] = J^{ss}_{\text{ net } A \rightarrow B} \times
\big(\mu_A-\mu_B\big).
\label{epr}
\end{equation}
We see that the steady state is actually an equilibrium if and only
if when $\mu_A=\mu_B$.  In this case, $\sigma[x^{ss}]$ is zero, 
$\lambda=0$, and one can check that $J^{ss}_{\text{ net } A \rightarrow B}=0$: There is no net transformation of $A$ to $B$
via $X$.  Otherwise, the $J^{ss}_{\text{ net } A \rightarrow B}\neq 0$, and it has the same sign as $\lambda$ and 
$(\mu_A-\mu_B)$.  Therefore,
the macroscopic entropy production rate defined
in (\ref{epr}),
which is never negative, mathematically quantifies the
statistical irreversibility in the self-replication of $X$.

Generalizing this example is straightforward; hence
with this in mind one can claim that \cite{jle-1}
``[e]very species of living thing can make a copy of itself
by exchanging energy and matter with its surroundings'',
which can be exactly computed if all the reversible
biochemical reactions involved are known.

\subsection{Time-dependent entropy production rate}

    We now turn our attention to the non-stationary
transient thermodynamics of system (\ref{the-rxn}).
First, the time-dependent concentration of $X$, $x(t)$
is readily solved from (\ref{the-ode}):
\begin{equation}
   x(t) = \frac{\beta C e^{-\beta t}}{1-Ck_{-1}e^{-\beta t}},
\end{equation}
in which
\[
        \beta = \sqrt{\big(k_{+1}a-k_{+2}\big)^2+4k_{-1}k_{-2}b },
           \   \
      C = \frac{x(0)}{k_{-1}x(0)+\beta}.
\]
Then, the time-dependent entropy production rate
$\sigma[x(t)]$, in $k_BT$ unit,
\begin{eqnarray}
    \sigma\big[x(t)\big] &=&
   \Big(k_{+1}ax(t)-k_{-1}x^2(t)\Big)
              \ln\left(\frac{k_{+1}a}{k_{-1}x(t)}\right)
\nonumber\\
    &&
            +\Big(k_{+2}x(t)-k_{-2}b\Big)
                  \ln\left(\frac{k_{+2}x(t)}{k_{-2}b}\right)
\\
    &=&   \textrm{cmf}\big[x\big]
              + \frac{\rd}{\rd t}\varphi^{ss}\big[x(t)\big].
\label{dgdt}
\end{eqnarray}
in which (see Eq. \ref{varphiss} below)
\begin{equation}
     \varphi^{ss}\big[x\big] = \int_0^x \ln\left(
           \frac{k_{-1}z^2+k_{+2}z }{
                  k_{+1}az+k_{-2}b}\right) \rd z ,
\label{gibbsfe}
\end{equation}
and the instantaneous chemical motive force
\begin{eqnarray}
   \textrm{cmf}\big[x\big] &=& \sum_{\ell=1,2}
             \Big(J_{\ell}^+(x)-J_{\ell}^-(x)\Big)
\nonumber\\
    &\times& \ln\left(\frac{J_{\ell}^+(x)}{J_{\ell}^-(x)}
               e^{-\nu_{\ell}\partial\varphi^{ss}(x)/\partial x}  \right)
\\
    &=&   \big(k_{+1}ax-k_{-1}x^2\big)
         \ln\left(\frac{k_{+1}ax}{k_{-1}x^2}
               e^{\partial\varphi^{ss}(x)/\partial x}  \right)
\nonumber\\
    &+&  \big(k_{+2}x-k_{-2}b\big)
         \ln\left(\frac{k_{+2}x}{k_{-2}b}
               e^{-\partial\varphi^{ss}(x)/\partial x}  \right),
\end{eqnarray}
with
\begin{equation}
       \frac{\partial\varphi^{ss}(x)}{\partial x} =  \ln\left(
           \frac{k_{-1}x^2+k_{+2}x }{
                  k_{+1}ax+k_{-2}b}\right).
\end{equation}
That is,
\begin{eqnarray}
  \text{cmf}\big[x\big] &=&  \big(k_{+1}ax-k_{-1}x^2\big)
\nonumber\\
    &\times&     \ln\left( \left(1+\frac{k_{+2}x}{k_{-1}x^2}\right) \left(1+\frac{k_{-2}b}{k_{+1}ax}\right)^{-1} \right)
\nonumber\\
    &+& \big(k_{+2}x-k_{-2}b\big)
\\
      &\times&   \ln\left( \left(\frac{k_{+1}ax}{k_{-2}b}+1\right)\left(\frac{k_{-1}x^2}{k_{+2}x}+1\right)^{-1}
              \right).
\nonumber
\end{eqnarray}
This is an example of Eq. \ref{dphixdt}.  Finally,
\begin{eqnarray}
    \frac{\rd}{\rd t}\varphi\big[x(t)\big] &=&
      \frac{\rd x(t)}{\rd t} \ln\left(
           \frac{k_{-1}x^2+k_{+2}x }{
                  k_{+1}ax+k_{-2}b}\right)
\nonumber\\
    &=& \Big(
      k_{+1}ax - k_{-1}x^2 -
                    k_{+2}x + k_{-2}b\Big) 
\nonumber\\
    &\times& \ln\left(
           \frac{k_{-1}x^2+k_{+2}x }{
                  k_{+1}ax+k_{-2}b}\right)  \le 0.
\label{dphixtdt}
\end{eqnarray}
This inequality implies even for a driven chemical 
reaction system that approaches to a NESS, there exists 
a meaningful ``potential function'' $\varphi[x]$ which
never increases. 

	To connect to the known Gibbsian equilibrium chemical
thermodynamics, we notice that chemical detailed balance
cmf$[x]=0$ $\forall x$ implies
\begin{equation}
            \frac{J_1^+(x)}{J^-_1(x)} =
      e^{\partial\varphi(x)/\partial x} =  \frac{J_2^-(x)}{J^+_2(x)},
\end{equation}
that is $k_{+1}k_{+2}a/k_{-1}k_{-2}b=1$.  Under this condition,
$\varphi[x]$ in (\ref{gibbsfe}) becomes
\begin{equation}
  \varphi\big[x\big] =  x\ln\left(
           \frac{k_{-1}x}{ k_{+1}a
          }\right) = x\big( \mu_X-\mu_A\big),
\end{equation}
which is the Gibbs function, in unit of $k_BT$, with state $A$
as reference.

The function $\varphi[x]$ in (\ref{gibbsfe}) is a nonequilibrium
generalization of the Gibbs free energy for open chemical
systems that approach to NESS \cite{ge-qian-2016,ge-qian-2016-2},
and Eq. \ref{dphixtdt} is an open-system generalization of the 
2nd Law.

\section{Stochastic Kinetic Description by the
Delbr\"uck-Gillespie Process}
\label{sec:dgp}

Chemical reactions at the individual molecule level are
stochastic \cite{wemoerner}, which can be described by the theory of
Chemical Master Equation (CME) \cite{erdi-book}
first appeared in the work of
Leontovich \cite{leontovich-35} and Delbr\"uck \cite{delbruck},
whose fluctuating trajectories can be exactly computed using the
stochastic simulation method widely known as Gillespie
algorithm \cite{gillespie}.  These two descriptions are
not two different theories, rather they are the two aspects of
a same Markov process, just as the Fokker-Planck equation
and the It\={o} integral descriptions of a same Langevin
dynamics.  More importantly, this probabilistic description and
the deterministic mass-action kinetics in Eq. \ref{the-ode}
are just two parts of a same
dynamic theory: The latter is the limit of the former if
fluctuations are sufficiently small, when the volume of
the reaction vessel, $\omega$, is large \cite{beard-qian-book}.

We now show this theory provides a more complete kinetic characterization of (\ref{the-rxn}) and it is in perfect agreement
with the classical chemical kinetics as well as Gibbs' chemical
thermodynamics.  Since chemical species in (\ref{the-rxn}) are discrete
entities, and the chemical reactions at the level of
single-molecules are stochastic, let $p_n(t)$ be the
probability of having $n$ number of $X$ in the reaction
system at time $t$.  Then, $p_n(t)$ follows the CME:
\begin{subequations}
\label{the-cme}
\begin{eqnarray}
   && \frac{\rd}{\rd t} p_n(t) =
               u_{n-1}p_{n-1} -\big( u_n+w_n\big) p_n
                 + w_{n+1}p_{n+1},
\nonumber\\[-5pt]
\\
    &&  u_n = u_n^{(1)}+u_n^{(2)},  \  \
          w_n =  w_n^{(1)}+w_n^{(2)}, \  \
          u_n^{(1)} = ak_{+1}n,
\nonumber\\[-3pt]
\\
  &&   u_n^{(2)} =  k_{-2}\omega b, \  \
        w_n^{(1)} = \frac{k_{-1}n(n-1)}{\omega}, \  \
        w^{(2)}_n = k_{+2}n.
\nonumber\\[-3pt]
\end{eqnarray}
\end{subequations}

\subsection{Stationary distribution}

The steady state of such an open chemical reaction
system has a probability distribution $\pi_n$ that is
no longer changing with time, after a certain period of
relaxation kinetics.  By open system, we mean it
has a constant chemical flux between $A$ and $B$,
with its direction being determined by which of the
$\mu_A$ and $\mu_B$ being greater.
The stationary probability distribution for $\pi_n$
is
\begin{eqnarray}
  \pi_n &=& C\prod_{\ell=1}^{n}
           \frac{ak_{+1}(\ell-1)+k_{-2}\omega b}{
                   k_{-1}\omega^{-1}(\ell-1)\ell+k_{+2}\ell }
\\
              &=&  \frac{C}{n!}
                \left(\frac{k_{-2}\omega b}{k_{+2}}\right)^n
               \left\{  \prod_{\ell=1}^{n}
           \frac{(\ell-1) e^{\frac{\mu_A-\mu_B}{k_BT}}
             + \left(\frac{\omega k_{+2}}{k_{-1}}\right) }{
                   (\ell-1)+\left(\frac{\omega k_{+2}}{k_{-1}}\right)}
                \right\},
\nonumber
\end{eqnarray}
in which $C$ is a normalization constant.  We observe that when
$\mu_A=\mu_B$, the $p_n^{ss}$ is a Poisson distribution,
as predicted by Gibbs' equilibrium theory of
grand canonical ensemble, with the mean number of
$X$ being $(k_{-2}\omega b/k_{+2})$, or equivalently
the mean concentration $(k_{-2}b/k_{+2})$.

The chemical thermodynamics presented above does not
reference to anything with probability.  But we know that the
very notion of Gibbs' chemical potential has a deep root in
it.

\subsection{Macroscopic limit}

With increasing $\omega$, the behavior of the Markov process
described by Eq. \ref{the-cme}
becomes very close to that described by the mass-action
kinetics.  In fact, if we let $x=n/\omega$ as the concentration
of the species $X$, then
\begin{equation}
     \frac{ p_{\omega x}(t) }{\omega} \sim  e^{-\omega\varphi(x,t)}
          \rightarrow \delta\big(x-z(t)\big),
\end{equation}
where
\begin{equation}
 \frac{\rd z(t)}{\rd t} = J^+(z) - J^-(z),
\end{equation}
\begin{equation}
   J^+(z) = \lim_{\omega\rightarrow\infty} \left(\frac{u_{\omega z} }{\omega}\right),
\   J^-(z) = \lim_{\omega\rightarrow\infty} \left(\frac{w_{\omega z} }{\omega}\right),
\end{equation}
and \cite{ge-qian-2016,hugang,shwartz-weiss-book}
\begin{eqnarray}
  \frac{\partial\varphi(x,t)}{\partial t} &=& \sum_{\ell=1}^2
           J^+_{\ell}(x) \left[1-e^{\nu_{\ell}\varphi'_x(x,t)}  \right]
\nonumber\\
        &+& J^-_{\ell}(x) \left[1-e^{-\nu_{\ell}\varphi'_x(x,t)]} \right].
\label{eqn16}
\end{eqnarray}
It turns out that the macroscopic, nonequilibrium chemical 
energy function appeared in Eq. \ref{dphixtdt} is the
stationary solution to Eq. \ref{eqn16}:
\begin{eqnarray}
   \varphi^{ss}(x) &=& -\int_0^x \ln\left(
           \frac{J_1^+(z)+J_2^-(z)
                 }{J_1^-(z)+J_2^+(z)}\right) \rd z,
\label{varphiss}
\end{eqnarray}
and
\begin{eqnarray}
    && \left(\frac{\partial\varphi^{ss}}{\partial x}\right)
\nonumber\\
            &=&  -\ln\left(
           \frac{J_1^+(x)+J_2^-(x)
                 }{J_1^-(x)+J_2^+(x)}\right)
\nonumber\\
    &=& -\ln\left(
           \frac{ ak_{+1}x+k_{-2}b
                 }{k_{-1}x^2+k_{+2}x}\right)
\label{eq37}\\
    &=& -\ln\left(
           \frac{k_{+2} \ e^{(\mu_B-\mu_X)/k_BT} + k_{-1}x
                          \ e^{(\mu_A-\mu_X)/k_BT}
                 }{ k_{+2} + k_{-1}x } \right),
\nonumber
\end{eqnarray}
in which $\mu_A-\mu_B = k_BT\ln\left(\tfrac{a k_{+1}k_{+2}}{bk_{-1}k_{-2}}\right)$.   We observe that when the reaction 
system is not driven chemically, $\mu_B=\mu_A$.  Then
(\ref{eq37}) is $\tfrac{\mu_X-\mu_A}{k_BT}$, the chemical
potential of $X$ in unit of $k_BT$, with state $A$
as reference.

\subsection{The stochastic NESS entropy production rate}

    According to stochastic thermodynamics, the NESS entropy
production rate for the stochastic dynamics in (\ref{the-cme}),
in $k_BT$ unit, is given in Eq. \ref{equation007}c:
\begin{eqnarray}
    e_p\big[\pi_n\big] &=& \sum_{n=0}^{\infty} \Big(\pi_nu_n^{(1)}-\pi_{n+1}
               w^{(1)}_{n+1}
               \Big)\ln\left(\frac{\pi_nu_n^{(1)}}{\pi_{n+1}
               w^{(1)}_{n+1}}\right)
\nonumber\\
    &&+\sum_{n=0}^{\infty} \Big(\pi_nu_n^{(2)}-\pi_{n+1}
               w^{(2)}_{n+1}
               \Big)\ln\left(\frac{\pi_nu_n^{(2)}}{\pi_{n+1}
               w^{(2)}_{n+1}}\right)
\nonumber\\
    &=& \sum_{n=0}^{\infty} \Big(\pi_nu_n^{(2)}-\pi_{n+1}
               w^{(2)}_{n+1}
               \Big)\ln\left(\frac{u_n^{(2)}w^{(1)}_{n+1}}{
                    w^{(2)}_{n+1}u_n^{(1)}}\right)
\nonumber\\
    &=& \sum_{n=0}^{\infty} \Big(\pi_nu_n^{(2)}-\pi_{n+1}
               w^{(2)}_{n+1}
               \Big)\ln\left(\frac{bk_{-1}k_{-2}}{
                          ak_{+1}k_{+2}}\right)
\nonumber\\
    &=& \omega\ln\left(\frac{ak_{+1}k_{+2}}{bk_{-1}k_{-2}}\right)
            \left(k_{+2}\frac{\langle n_X\rangle^{ss}}{\omega} -k_{-2}b\right).
\label{s-epr}
\end{eqnarray}
This agrees exactly with Eq. \ref{epr}, which is the entropy
production rate per unit volume.

\section{Coarse Graining: Kinetics and Thermodynamics}
\label{sec:KandT}

    The entropy production given in (\ref{epr}) and
(\ref{s-epr}) can be decomposed into two
terms:
\begin{equation}
    \sigma[x^{ss}]
      =  \Big\{ J^{ss}_{ \text{net } A\rightarrow X } \times
                \big(\mu_A-\mu_X\big) \Big\}
             + \Big\{ J^{ss}_{\text{ net } X \rightarrow B} \times
                \big(\mu_X-\mu_B\big) \Big\},
\end{equation}
in which $J^{ss}_{ \text{net } A\rightarrow X}=$
$k_{+1}ax^{ss}-k_{-1}(x^{ss})^2$,
$\mu_A-\mu_X=k_BT\ln\big(k_{+1}a/(k_{-1}x^{ss})\big)$,
$J^{ss}_{\text{ net } X \rightarrow B}=$
$k_{+2}x^{ss}-k_{-2}b$, and
$\mu_X-\mu_B=k_BT\ln\big(k_{+2}x^{ss}/(k_{-2}b)\big)$.
Both terms in $\{\cdots\}$ are positive.  More importantly,
if $k_{+1}$ and $k_{-1}$ are very large while
$J^{ss}_{\text{net } A\rightarrow X}$ is kept constant, then
$\mu_A-\mu_X$ will be very small; they are nearly at
equilibrium.   In this case, we can
lump the $A$ and $X$ as a single
chemical species with rapid internal equilibrium.  Such
coarse-graining always leads to under-estimating the entropy
production \cite{santillian-qian,esposito}.

\subsection{Thermodynamically consistent kinetic
approximation}
\label{sec:thermoconsistent}

    We now carry out a more in depth analysis on
kinetic coarse-graining.
In particular, we try to show the following: There are at least
two types of kinetic approximations: those are
thermodynamically meaningful and those are not.
In mathematical terms: No matter how inaccurate, the former
gives a finite approximation of the ``true entropy production of
the system'' while the latter yields a numerical infinity,
{\em e.g.}, the thermodynamics is lost.

    According to T. L. Hill \cite{tl-hill-book}, 
the expression in (\ref{nessflux})
has a unique, thermodynamically meaningful representation
$J^{ss}_{\text{net } A \rightarrow B} =$
$J^{ss}_{A \rightarrow B}-J^{ss}_{B \rightarrow A}$,  in which $J^{ss}_{A \rightarrow B}=ak_{+1}k_{+2}/\Sigma$,
$J^{ss}_{B \rightarrow A}=bk_{-1}k_{-2}/\Sigma$, and
\begin{equation}
        \Sigma = \frac{1}{k_{-1}}
               +\left(\frac{k_{+2}}{2k_{-1}}\right)
      \frac{ \sqrt{\big(k_{+1}a+k_{+2}\big)^2
                  -4\lambda}-
           \big( k_{+1}a+k_{+2}\big) }{\lambda}.
\end{equation}
One can show that $\Sigma$ is strictly positive if
all the parameters having finite values.  Then,
the entropy production rate in (\ref{epr})
\begin{equation}
     \sigma\big[x^{ss}\big]
      =\Big(J^{ss}_{A \rightarrow B}-J^{ss}_{B \rightarrow A}\Big)
                \ln\left(\frac{J^{ss}_{A \rightarrow B}}{J^{ss}_{B \rightarrow A}}
                    \right).
\label{epr2}
\end{equation}

    With respect to the expression in Eq. \ref{epr2},
we can reach the following conclusions:

$(a)$
If an approximation leads to $J^{ss}_{A\rightarrow B}\rightarrow
J^*_+$, $J^{ss}_{B \rightarrow A}\rightarrow J^*_-$ but
$J^*_+/J^*_-$ finite, then there is a meaningful,
finite approximated $\sigma$.  This is a sufficient
condition but not necessary.

$(b)$ If both $J^{ss}_{A \rightarrow B}$,
$J^{ss}_{B \rightarrow A}\rightarrow\infty$,
but $(J^{ss}_{A\rightarrow B}-J^{ss}_{B \rightarrow A})$
is finite, then there is a
finite $\sigma$.

$(c)$ It is possible that
$(J^{ss}_{A \rightarrow B}-J^{ss}_{B\rightarrow A})\rightarrow\infty$, $(J^{ss}_{A \rightarrow B}/J^{ss}_{B\rightarrow A})\rightarrow 1$,
and $\sigma$ is finite.
For example, $J^{ss}_{A \rightarrow B}= x+x^2$,
$J^{ss}_{B\rightarrow A}=x^2$.
Then when $x\rightarrow\infty$ we have
$\sigma=1$.

$(d)$ However, if one of the $J^{ss}_{A \rightarrow B}$
and $J^{ss}_{B\rightarrow A}$ is finite and another one
is not, then $\sigma=\infty$.

    We see that if $k_{-1},k_{-2}\rightarrow 0$ while
all other parameters are finite, then it is the scenario $(d)$.
In this case, a meaningful thermodynamics no longer
exists for the kinetic approximation.

\subsection{Birth-and-death approximation of replication
kinetics}

    For a macroscopic sized system (\ref{the-rxn})
with externally sustained $a$ and $b$,
the equation for its chemical kinetics is
\begin{equation}
  \frac{\rd x}{\rd t} =  \big(k_{+1}a+k_{-2}b\big) x
              - \big(k_{-1}x+k_{+2}\big) x.
\label{lma-1}
\end{equation}
Let us now consider a particular case in which
\begin{equation}
   k_{+1}a \gg k_{-2}b  \  \textrm{ and  } \
    x\ll k_{+2}/k_{-1}.
\label{k-app}
\end{equation}
Then it is completely legitimate to approximate the
kinetic equation in (\ref{lma-1}) by an approximated
\begin{equation}
  \frac{\rd x}{\rd t} = \big(g- k_{+2}\big) x, \  \
                    g\equiv k_{+1}a.
\label{lma-2}
\end{equation}
In fact, the approximation produces accurate
kinetic result if the two inequalities in (\ref{k-app})
are strong enough.

    If system (\ref{the-rxn}) is mesoscopic sized, then
the deterministic dynamics in (\ref{lma-2}) also has a
stochastic, Markov counterpart, as a {\bf\em birth-and-death}
process with master equation for the probability distribution
$p_n(t)\equiv\Pr\big\{n_X(t)=n\big\}$, where $n_X(t)$ is the
number of $X$ in the system at time $t$:
\begin{eqnarray}
      \frac{\rd}{\rd t}p_n(t) &=& g\Big((n-1)p_{n-1}-np_n\Big)
\nonumber\\
           &-& k_{+2}\Big( np_n(t) - (n+1)p_{n+1}(t)\Big).
\label{bdp}
\end{eqnarray}
For very large $n$, one can approximate $n+1\approx n\approx n-1$,
then this is the equation [9] in ref. \cite{jle-1}, if we identify
$k_{+2}$ as $\delta$.

    However, as discussed in Sec. \ref{sec:thermoconsistent},
while coarse-graining is a type of mathematical
approximation, not all mathematically legitimate kinetic
approximations are valid coarse-graining for thermodynamics.
The approximated (\ref{lma-2}) now yields meaningless
chemical thermodynamics.  To see this, we consider the exact
result in Eq. \ref{dmu}
\begin{equation}
  \mu_{A}-\mu_B = \ln\left(\frac{ak_{+1}k_{+2}}{
                 bk_{-1}k_{-2}}\right),
\label{chemep}
\end{equation}
and compare it with the mathematical entropy production
in the Markov transition of the birth-and-death
process (\ref{bdp}), from $n_X=n$ to $n_X=(n+1)$:
\[
      \text{entropy production of birth-and-death}
\]
\[
             = \ln \left( \frac{\Pr\{ \text{a birth in $\Delta t$ time} \} }{
                         \Pr\{ \text{a death in $\Delta t$ time}  \} }
                 \right)
\]
\begin{equation}
    = \ln\left(\frac{ank_{+1}p_n}{k_{+2}(n+1)p_{n+1}}\right).
\label{bdpep}
\end{equation}
Comparing Eq. \ref{bdpep} and Eq. \ref{chemep},
we see there is no definitive relation, nor definitive inequality
between the two quantities: Both $k_{+1}$ and $k_{+2}$,
which are on the numerator and denominator in
(\ref{bdpep}), are on the numerator in (\ref{chemep}).

    One also notices that when $k_{-1}=0$, the Markov process
described by Eq. \ref{the-cme} is possible to continously
increase without reaching stationarity:
\begin{equation}
           \Big(ak_{+1}n+k_{-2}b\omega \Big)
          > k_{+2}n,  \   \   \forall n.
\label{eqn0047}
\end{equation}
However, for $k_{-1}\neq 0$, the inequality in (\ref{eqn0047})
can not be valid for all $n$: There must be an $n^*$, when
$n>n^*$:
\begin{equation}
           \Big(ak_{+1}n+k_{-2}b\omega \Big)
          < \left(\frac{k_{-1}(n-1)}{\omega}+k_{+2} \right)n.
\end{equation}
This shows that  a chemical kinetics with meaningful thermodynamics is
intrinsically stable.

    As we have stated earlier, considering microscopic reversibility
is a fundamental tenet of stochastic thermodynamics.
Any coarse-graining that is {\bf\em thermodynamically meaningful}
has to respect the nature of microscopic reversibility.
On the other hand, the death step represented by $\delta=k_{+2}$
simply is not the reverse step of the birth step represented by
$g=k_{+1}a$;  it cannot provide any meaningful estimation.
The reversed step of birth is actually an infant
going back to the mother's womb!  In a population dynamics
model like (\ref{lma-2}), while it could be completely
accurate in kinetic modeling, nevertheless is not thermodynamically
meaningful on the level of physical chemistry.  There are
irreversible approximations involved.

    As pointed out in \cite{jle-1}, ``it is much more likely that one bacterium should turn into two than that two should somehow spontaneously revert back into one.''  Nevertheless, it is the
non-zero probability of such completely inconceivable processes,
in a physicist's world, that provides an estimation of thermodynamic
heat production.  When a biologist considers only feasible
events to a living organism being birth and death, the connection
to the physical world has already lost in the mathematical
thinking.

\section{Conclusion and Discussion}
\label{sec:dis}

{\bf\em Another 2nd Law for systems with irreversible
processes?}
First, we should emphasize that while classical mechanics
represents a system in terms of point masses each with
a distinct label, reasonable ``variables'' in biology,
physiology, and biochemistry, more often are counting numbers
of various ``species'':\footnote{Note that when dealing with
many-body systems, fluid mechanics and quantum mechanics
changed their representations from tracking the state of
individually labeled particles to counting the numbers:
the switching from Lagrangian to Eulerian descriptions in
the former and second quantization in the latter.
} number of molecules in biochemistry,
number of cells in a tissue, and number of individuals in a
population.  In these latter cases, the Crooks' equality, e.g., Eq. (3) in
\cite{jle-1} is mathematically hold, but its interpretation
as ``heat'', based on Gibbs' chemical thermodynamics,
is subtle \cite{ge-qian-2013}: It depends on how the environment
of an open system is set up.  The discussion in
Sec. \ref{sec:KandT} serves as a warning. Note when
counting the number of individuals in a population, say $n_X$,
the increase of $n_X$ from $n$ to $n+1$, and
its decrease from $n+1$ to $n$, even they are through
reversible reaction steps, it cannot be identified whether
they occur by a same individual or different individuals:  This
information has already lost in the number counting representation 
of nameless individuals.  However, since the mathematics is valid, 
it is legitimate, even scientifically desirable, to propose a different
type of ``heat'' in Markov systems that is not necessarily connected 
to mechanical energy via $k_BT$.

Also, with a thermodynamically valid coarse-graining,
the computation of entropy production on a coarse level
can provide a lower bound for the {\em mechanics based
free energy} dissipation, as stated in \cite{jle-1}.  This result is
not new; it has been discussed already
in \cite{santillian-qian} and \cite{esposito}.  In a nutshell,
coarse-graining involves ``rapid equilibrium'' assumption
which hides dissipation in the fast modes of motions.
In fact, with a given difference $\big(J^+-J^-\big)$
which fixes the rate of an irreversible process,
the entropy production $\ln\big(J^+/J^-\big)$
decrease with increasing one-way fluxes $J^+$ and
$J^-$.

{\bf\em The physics of living systems is outside classical
mechanics.}  While no living phenomenon and process
disobey classical mechanics, the former are phenomena
with such a large degree of freedom, heterogeneity, and
complexity, their understanding has to be founded on
laws and descriptions outside classical
mechanics. By classical mechanics, we mean the view of
the world in terms of point masses and their movement
based on the Newtonian system of
equations of motion.  In particular, we try to show that
the understanding of ``heat'', which is so fundamental
in the irreversibility of macroscopic mechanical systems, actually
has very little to do with irreversibility in living systems!
In contrast, chemical thermodynamics and its irreversibility,
{\em \`{a} la} Gibbs, Lewis and Randall,
naturally arises from the theory of probability, the mathematics
J. W. Gibbs employed in terms of the notion of {\em ensemble},
for developing his brand of {\em statistical mechanics}.  It did
not escape our notice that Gibbs was also the originator of the
chemical thermodynamics, and the inventor of the notion of
chemical potential.

    Understanding thermodynamics from Newtonian mechanics
was the central thesis of L. Boltzmann's life work.  Indeed,
one of the key results of Boltzmann, together with H. von
Helmholtz, was to cast the phenomenological First Law of thermodynamics
into a mathematical corollary of the Newtonian equation
of motion:  One knows a Hamiltonian dynamics is restricted
on a level set of $H(\{x_i\},\{v_i\})=E$, where the $E$
is determined by the initial condition.   Recognizing that a
thermodynamic state is actually a state of perpetual motion,
e.g., an entire level
set\footnote{Here, the importance of {\em ergodicity} arises.},
there is a mathematical function between the phase-space
volume contained by the level-set $\Omega$, $E$, and other
parameters in the Hamiltonian function $H(\{x_i\},\{v_i\},V,N)$:
$\Omega(E,V,N)$.  Since the phase-space volume monotonically increases
with $E$, one can define
\begin{equation}
      E = E(S,V,N), \   \textrm{ where }  S=\ln\Omega(E,V,N).
\end{equation}
It is then a matter of simple calculus to obtain
\begin{equation}
  \rd E = \left(\frac{\partial E}{\partial S}\right)_{V,N}\rd S
          + \left(\frac{\partial E}{\partial V}\right)_{S,N}\rd V
          + \left(\frac{\partial E}{\partial N}\right)_{S,V}\rd N,
\end{equation}
and define the emergent thermodynamic quantities
$T=\big(\partial E/\partial S\big)_{V,N}$ as temperature,
$p=-\big(\partial E/\partial V\big)_{S,N}$ as pressure,
and
$\mu=\big(\partial E/\partial N\big)_{S,V}$ as chemical
potential.  Note, from the mechanical standpoint,
$T,p,\mu$ are emergent quantities; they characterize
how one invariant torus is related to another invariant torus,
i.e., one thermodynamic equilibrium state to another
thermodynamic equilibrium state.

    It is important to emphasize that, no matter how
imperfect, both $T$ and $p$ have well-accepted and
widely understood mechanical interpretation,
as mean kinetic energy and mean momentum transfer to
the wall of a container, respectively.
However, $\mu$ has no interpretation in
terms of classical motion, whatsoever; rather, it has an interpretation
in terms of probability, and in terms of Brownian motion:
\begin{equation}
  \frac{\partial \rho(x,t)}{\partial t}
      = D\frac{\partial^2\rho(x,t)}{\partial x^2}
      = -\frac{1}{\eta}\frac{\partial (\hat{F}\rho )}{\partial x},
\end{equation}
where
\begin{equation}
       \hat{F} = -\frac{\partial\mu}{\partial x},  \
            \textrm{ and } \
       \mu = D\eta\ln\rho(x,t) = k_BT\ln\rho(x,t).
\end{equation}
$\hat{F}$ is known as {\em entropic force} in chemistry,
and $\mu$ is its potential function.

    A living system is sustained neither by a
temperature difference, nor by a pressure difference.  It is
a phenomenon driven by chemical potential difference.
Therefore, we believe any discussion of irreversibility in
living systems based on the notion of heat is misguided.
Nevertheless, as the physicist in 18$^{th}$ century had recognized
in connection to the notion of ``heat death of the universe'',
a sustained chemical potential difference has to have a
consequence in generating heat in a closed, cyclic universe,
if one indeed can treat the entire universe as an isolated
mechanical system.  However, it is equally likely, according
to our current, limited understanding of the cosmology
and planet formation, that what is being dissipated is
simply local inhomogeneity, e.g., low entropy initial condition,
in our world, which was formed 13 some billion years ago.

{\bf\em Information and entropy.}
There is a growing interest in the phenomenon of
``information to energy conversion'' in Maxwell-demon
like devices  \cite{sano-2010}.  To quantify information,
entropy change has to be distinguished from work.
An unambiguous distinction between work and heat
is fundamental to thermodynamics: ``{\it The difference
between heat and work cannot always be uniquely
specified. It is assumed that there are cases, involving
ideal processes, in which the two can be strictly
distinguished from one another}'' \cite{pauli-book}.
Unfortunately outside classical mechanics, at least in
biochemistry where temperature-dependent heat
capacity is a commonplace \cite{benzinger-70},
distinguishing enthalpy change from entropy change
becomes increasingly challenging
\cite{qian-jjh-96,qian-jcp-98,ge-qian-2013}.
In a more modern physics setting, heat, probability, and 
time are intimately related \cite{rovelli-carlo-book}.
The success of our present theory on chemical thermodynamics
owes to a large extent to a sidestepping the notion of
heat, and thus temperature that follows.

{\bf\em The physical locus of entropy production.} Whether
the entropy is produced inside a subsystem, or in the environment
outside the system, or just at the boundary \cite{sekulic-01},
is one of the deeper
issues in nonequilibrium thermodynamics:  To L. Onsager, a transport process driven by a {\em thermodynamic force} constitutes dissipation \cite{onsager-31}. This is a NESS view of an ``insider'' of an open subsystem; it can be mathematically justified in terms of the positive
chemical motive force (cmf) introduced in Eq. \ref{cmf}.
To classical physicists, however, the mechanistic origin of the
``thermodynamic force'' is outside the subsystem due to
spontaneous processes that cause entropy of the total closed
system to increase.  This view is justified in terms of
$\tfrac{\rd F}{\rd t}\le 0$ for closed systems.  We refer the
readers to T. L. Hill's notion of ``cycle completion''
in stochastic thermodynamics \cite{hill-pnas-83} for a much more
complete view of the matter.  This concept nicely
echoes R. Landauer and C. H. Bennett's principle of computational
irreversibility being associated only with memory erasing
\cite{landauer,bennett}.  It also provides a powerful conceptual
framework for further investigating other nagging concepts
such as heat dissipation associated with a subsystem in
a nonequilibrium steady state (NESS) \cite{ge-qian-2013}
and endo-reversibility in finite time thermodynamics
(FTT) \cite{sekulic-01,qian-epjst-15a}.

 

\section*{Acknowledgment}

D. B. S. and H. Q. acknowledge the MOST grant 104-2811-M-001-102 
and the NIH grant R01-GM109964, respectively, for support.  
H. Q. thanks Ken Dill, Tolya Kolomeisky, and Chao Tang for pointing 
him to the problem of self-replication, that initiated the present 
work.  

\ifCLASSOPTIONcaptionsoff
  \newpage
\fi



%

%

\begin{IEEEbiographynophoto}{David B. Saakian}
is a lead researcher of A. I. Alikhanyan
National Science Laboratory (Yerevan Physics Institute)
Foundation and a visiting scholar at the Institute of Physics, 
Academia Sinica.
He graduated from Moscow Engineering Physics Institute in
1977 and obtained his Ph.D. in physics  from Moscow Lebedev
Institute of Physics of Russian Academy of Science in 1980. 
He had worked in the fields of optics and high energy physics
before moved to Yerevan Physics Institute and 
started research in statistical physics of disordered systems. 
In 1990s, the group he led developed an
optimal coding theory using the Random Energy
Model of Derrida.  His current research is focused
on the solutions of evolution models, for which he had
developed methods for obtaining exact results.  These
include the dynamics of Crow-Kimura and Eigen
models.  Very recently he derived exact results for
solutions of chemical master equations (CMEs)
by organizing them as a strip of 1-dimensional chain
of equations.  He has served as referees for numerous 
top journals in the field of evolution models and CMEs.
\end{IEEEbiographynophoto}

\begin{IEEEbiographynophoto}{Hong Qian}
received his B.A. in astrophysics from Peking University
and obtained his Ph.D. in molecular biology from
Washington University (St. Louis). His research interests
turned to theoretical biophysical chemistry and mathematical biology
when he was a postdoctoral fellow at the University of Oregon
and at the California Institute of Technology, where
he studied protein folding.
He was with the Department of Biomathematics at 
the UCLA School of Medicine between 1994 and 1997, 
when he worked on the theory of motor proteins and 
single-molecule biophysics. He joined the University of 
Washington (Seattle) in 1997 and is now
professor of applied mathematics. 
He is a fellow of the American Physical Society,
and served and serves on the editorial board of journals on
computational and systems biology.  His current research
interest is in stochastic analysis and statistical physics of 
cellular systems.  He has coauthored a book 
``Chemical Biophysics: Quantitative Analysis of
Cellular Systems'' (Cambridge University Press, 2008) with Daniel A. Beard.

\end{IEEEbiographynophoto}





\end{document}